\documentclass[amsfonts, amssymb, amsmath, superscriptaddress, showpacs, reprint]{revtex4-1}
\usepackage{times}
\usepackage{bm}
\usepackage{hyperref}
\usepackage{graphicx}
\usepackage{epsfig} 

\def\etal{\mbox{\em et al.}}

\def\kk{\mathbf k}
\def\pp{\mathbf p}
\def\qq{\mathbf q}
\def\si{{\scriptstyle{K}}}
\def\sii{{\scriptstyle{P}}}
\def\siii{{\scriptstyle{Q}}}
\def\kp{p}
\def\kc{c}

\begin{document}

\title{Locality properties of the free energy fluxes in gyrokinetic turbulence}
\author{Bogdan Teaca}
\email{bogdan.teaca@epfl.ch, Tel: +41.21.693.4305}
\affiliation{Ecole Polytechnique F\'ed\'erale de Lausanne (EPFL), Centre de Recherches en Physique des Plasmas, Association Euratom-Conf\'ed\'eration Suisse, CH-1015 Lausanne, Switzerland.}
\author{Alejandro \surname{Ba\~n\'on Navarro}}
\affiliation{Statistical and Plasma Physics, Faculty of Sciences, Universit\'e Libre de Bruxelles, Campus Plaine, CP 231, B-1050 Brussels, Belgium.}
\author{Frank Jenko}  
\affiliation{Max-Planck-Institut f\"ur Plasmaphysik, EURATOM Association, 85748 Garching, Germany.}
\author{ Stephan Brunner}
\affiliation{Ecole Polytechnique F\'ed\'erale de Lausanne (EPFL), Centre de Recherches en Physique des Plasmas, Association Euratom-Conf\'ed\'eration Suisse, CH-1015 Lausanne, Switzerland.}
\author{Laurent Villard}
\affiliation{Ecole Polytechnique F\'ed\'erale de Lausanne (EPFL), Centre de Recherches en Physique des Plasmas, Association Euratom-Conf\'ed\'eration Suisse, CH-1015 Lausanne, Switzerland.}
\begin{abstract}
The nature of the nonlinear interactions in gyrokinetic (GK) turbulence, driven by an ion-temperature gradient instability, is investigated using numerical simulations of single ion species plasma in three-dimensional flux tube geometry. To account for the level of separation existing between scales involved in an energetic interaction, the degree of locality of the free energy scale flux is analyzed employing Kraichnan's infrared (IR) and ultraviolet locality functions. Due to the nontrivial dissipative nature of GK turbulence, an asymptotic level for the locality exponents, indicative of a universal dynamical regime for GK's, is not recovered and an accentuated non-local behavior of the IR interactions is found instead, in spite of the local energy cascade observed.
\end{abstract}
%
%
\maketitle

{\em Introduction.---} The main characteristic of physical flows is given by the existence of couplings between different scales of motion ($\ell$), described mathematically by nonlinear terms in respect to the dynamical quantities. As result of these couplings, the flow will develop a turbulent state for a sufficiently large interval of excited scales. While the range of scales available to the flow depends on the boundary conditions and on the nature of the flow itself (mathematically described by linear terms and externally given parameters), the redistribution of information among different scales is only due to the nonlinear terms. As such, the redistribution mechanism is expected to have a universal behavior for intervals of scales for which the linear terms are negligible, i.e. the inertial zone. This picture stands at the basis of the study of turbulence for hydrodynamical and electrically conductive fluids. In principle, gyrokinetic (GK) turbulence makes no exception to this picture \cite{Schekochihin:2008p1034, Tatsuno:2009p1096}, albeit with a series of complications due to the nature of the linear terms.

The gyrokinetic formalism is pertinent for the study of multi-species plasmas in the presence of strong magnetic guide fields \cite{Krommes:2012p1373}. By eliminating exactly the gyration phase of charged particles around the magnetic field lines \cite{Brizard:2007p11, Cary:2009p1231}, the dynamical space can be reduced from six dimensions to five. From the start, it is seen that the constraint imposed by the magnetic guide field on the charged flow creates an anisotropy in the system ($\ell=\{\ell_\perp, \ell_\parallel\}$). The scaling in the two directions are linked as result of causality, a hypothesis known as critical balance, recently used in the scaling of plasma turbulence \cite{Schekochihin:2009p1131, Barnes:2011p1372}. The current work concentrates on the analysis of the perpendicular spatial structures ($\ell_\perp$) of the fluctuations. 

To understand the dynamics introduced by the nonlinear term, the scale redistribution of free-energy (a GK ideal invariant, i.e. a global quantity that remains constant in time in the absence of source and sink effects) is usually investigated. Different works reported that the exchange of energy takes place between closest neighbor dyadic structures \cite{Plunk:2010p1360, BanonNavarro:2011p1274, Plunk:2011p1357}. However, although the energy exchanges are local, the question regarding the locality of the interactions was never addressed, i.e. the fact that the local exchanges of energy might be generated by the interaction of highly separated (non-local) scales. In this work, we describe a quantitative way of asserting the degree of locality for GK turbulence. The idea of locality can be seen as the disparity of scales contributing to a nonlinear interaction. For an given energy flux through a scale, the degree to which each scale contributes to the mentioned flux represents a useful assertion of locality of interactions, \cite{Kraichnan:1959p497, Teaca:2011p1362}. For the interaction to be local, the contribution of highly separated scales should be small and decrease fast with the increase in separation.

For fluid turbulence, the separation of the forced and dissipative scale ranges leads to the existence of a natural inertial range that possesses an unique locality exponent, which asymptotes dynamically to a $4/3$ value \cite{Zhou:1993p949}. However, in the case of GK turbulence and for fusion plasma configurations in particular, each instability that generates a time unstable mode (energy source) is also accompanied by an ensemble of time stable modes (energy sinks). These stable modes, known as linear damped eigenmodes, are nonlinearly coupled to the unstable modes and are responsible for an additional energy dissipation route \cite{Hatch:2011p1369}. This dissipation mechanism acts at a scale comparable to the forcing and as such, does not require the existence of a classical nonlinear cascade. Thus, the existence of a nonlinear cascade process is made possible only if the energy injected by the unstable modes is greater than the energy dissipated locally by the linear damped eigenmodes. Moreover, since the dissipation tends to permeate strongly the forced range and take over from the force as the dominant effect for smaller scales, characteristic time wise, the nonlinear cascade process occurs inside a dissipation range \cite{BanonNavarro:2011p1350, Terry:2009p1359}. This represents a novel and nontrivial complication compared to fluid turbulence, which leads to a dampening of contributions the to energy flux. As result, a unique locality exponent can not be found, in spite of the local energy exchanges.

{\em Gyrokinetic simulations.---}In the present work, numerical solutions of the nonlinear gyrokinetic equations in flux-tube ($\hat s - \alpha$) geometry \cite{Lapillonne:2009p1355} are analyzed using a field-aligned coordinate system ($x, y, z, v_\parallel, \mu$) with $(128,64,16,32,8)$ points in each direction, respectively. The solutions are obtained by the use of the {\sc gene} code, \cite{Jenko:2000p1248} for ion-temperature gradient (ITG) driven GK turbulence with physical parameters corresponding to the Cyclone Base Case \cite{Dimits:2000p1375}. For a better understanding of the nonlinear dynamics involved, the analysis is limited to the simple scenario of electrostatic fluctuations generated by a single ion species (the species index will be omitted) and adiabatic electrons, in the context of a large aspect-ratio circular cross-section equilibrium model, for which the equilibrium magnetic field is $B_0$ (in units of the magnetic field value on the magnetic axis). For details see Ref.~\cite{BanonNavarro:2011p1350}.

Considering that the total ion distribution function $F$ is split into an appropriately normalized Maxwellian part $F_{0}=\pi^{-3/2}e^{-(v_\parallel^2+\mu B_0)}$ and a perturbed part $f$, the non-adiabatic contribution of the ion distribution function is given as $h=f+(Z\bar\phi_1/T_{0})F_{0}$, where $\bar\phi_1$ is the gyro-averaged electrostatic potential, $T_{0}$ is the ion temperature (normalized to the electron temperature) and $Z$ is the electric charge. The time ($t$) evolution equation for the perturbed distribution function reads,
\begin{align}
\frac{\partial f}{\partial t}&=G[f]+L_C[f]+L_\parallel[f]+ N[f,f]+D[f]\label{GKeq} \;,
\end{align}
where $G[f]=-\left[\omega_{n}+\left( v_\parallel^2+\mu B_0-\frac{3}{2}\right) \omega_{T}\right] F_{0}\frac{\partial \bar\phi_1}{\partial y} $ is the contribution of the normalized background density ($\omega_{n}=-R\partial \log n_{0}/\partial x$) and temperature ($\omega_{T}=-R\partial \log T_{0}/\partial x$) gradients, with $R$ being the the major radius. It represents the driving mechanism for GK turbulence and it is responsible for the injection of free energy into the system. The second term appears due to magnetic curvature, $L_C[f]=\frac{T_{0}(2v_\parallel^2+\mu B_0)}{Z B_0}\left(2\sin z \frac{\partial h}{\partial x}+2(\cos z +\hat s z \sin z)\frac{\partial h}{\partial y} \right)$ as result of the $\hat s - \alpha$ geometry ($\alpha=0$) employed here. The third terms contains the parallel dynamics involving magnetic trapping and linear Landau damping/pumping effects, $L_\parallel[f]=-\frac{v_{T}}{2}\left[ v_\parallel^2+\mu B_0, h\right]_{zv_\parallel}$, where $v_{T}=\sqrt{2T_{0}/m}$ is the ion thermal velocity, $m$ is the ion mass and the Poisson bracket structure is defined as: $[f,g]_{ab}=\partial_a f\partial_b g-\partial_b f\partial_a g$. The second to last term represents the nonlinear term, $N[f,f]=-[\bar\phi_1,h]_{xy}$, while the last term in Eq.~(\ref{GKeq}) contains the dissipative effects. To avoid the computational cost of collisional operators, the dissipation terms have a simple hyper-diffusivity form $D[f]=-(a_z \partial^n_z + a_{v_{\parallel}} \partial^n_{v_{\parallel}}) f $, where $n=4$ and the $a$'s parameters are adapted to the problem at hand \cite{BanonNavarro:2011p1274, Hatch:2011p1369, BanonNavarro:2011p1350}. A $2/3$ dealiasing method is used for the nonlinear term. 

To close the equation system, the self-consistent electrostatic potential is obtained by solving the gyrokinetic Poisson equation,$\frac{Z^2 n_{0}}{T_{0}}[1-\Gamma_0(b)]\phi_1= Z\pi B_0n_{0}  \int J_0(\lambda) f \mbox{d}v_{\parallel} \mbox{d}\mu$, where the Bessel function  $J_0$, $\Gamma_0(b)=e^{b}I_0(b)$ and the modified Bessel function $I_0$ have the arguments defined as, $\lambda=\sqrt{\mu B_0}k_\perp v_{T}/\Omega$ and $b=k^2_\perp v^2_{T}/(2\Omega^2)$, with $\Omega=ZB_0/(m c)$ and $k_\perp$ being the perpendicular wavenumber.

{\em Free energy balance.---} In this formulation, the global free energy contained in the system is defined as, $\mathcal E =\frac{1}{2}\int \mbox{d}x \mbox{d}y \mbox{d}\Theta\ \frac{T_{0}}{F_{0}}\; h f\;,$ where $\mbox{d}\Theta=(\pi B_0 n_{0})\mbox{d}z\mbox{d}v_{\parallel}\mbox{d}\mu$. To analyzed the excitation degree of perpendicular turbulent scales, an integral over the $\mbox{d} \Theta$ infinitesimal element and a Fourier decomposition of the remaining $(x, y)$ space are performed. In the field aligned coordinates, the $x$ label refers to the flux surface, while $y$ identifies different field lines laying on the same flux surface. Each scales of length ($\ell_\perp$) can now be easily identified by the norm ($k\sim\ell_\perp^{-1}$) of the wave-vector ($\kk\equiv\kk_\perp$) based in the $k_x$, $k_y$ space. 

As for any quadratic quantity, the free energy spectral density can be considered ($\mathcal E=\sum_\kk \mathcal E^\kk$), for which the balance equation reads,%
\begin{align}
\partial_t \mathcal E^\kk &=  \mathcal T^\kk  +\mathcal L^\kk + \mathcal G^\kk + \mathcal D^\kk \;,\label{evoent_k}
\end{align}
where in the {\em rhs} of Eq.~(\ref{evoent_k}) the terms $\mathcal A^\kk=\{\mathcal G^\kk, \mathcal L^\kk,\mathcal D^\kk\}$ are found as, $\mathcal A^\kk= \int \mbox{d} \Theta\ \frac{T_{0}}{F_{0}} h^\kk A^{-\kk} $ using the spectral form of the evolution equation for the perturbed distribution function Eq.~(\ref{GKeq}), with $A^\kk=\{G^\kk,  L^\kk=L^\kk_C+L^\kk_\parallel, D^\kk\}$. While in Eq.~(\ref{evoent_k}) the linear quantities $\mathcal L$, $\mathcal D$ and $\mathcal G$ are defined involving only $\kk$ local modes and their complex conjugates, for the term generated by the nonlinear product $\mathcal T^\kk$ different modes enter in the definition,
\begin{align}
\mathcal T^\kk   &=  \sum_\pp \sum_\qq \mathcal T^{\kk, \pp, \qq}\delta_{\kk+\pp+\qq} \;.
\end{align}
The discreet Delta Dirac $\delta_{\kk+\pp+\qq}$ selects only interactions that occur between a triad of modes which obey the resonance condition, $\kk+\pp+\qq={\bf 0}$. The transfer that takes place for a single triad, known as the {\em triad transfer}, is defined as,
\begin{align}
\mathcal T^{\kk, \pp, \qq}=\int\ \mbox{d}\Theta \frac{T_{0}}{2F_{0}} &\Big{[} q_x p_y -q_y  p_x \Big{]} \Big{[} \bar \phi_1^{\qq} h^{\pp} -\bar \phi_1^\pp h^\qq \Big{]}  h^\kk \;, 
\end{align}
where the symmetry in modes $\qq$ and $\pp$ is written explicitly \cite{Nakata:2012p1387}. 
At the triad level, the free-energy conservation by the nonlinear interaction can be written as, $\mathcal T^{\kk, \pp, \qq}+\mathcal T^{\pp, \qq, \kk}+\mathcal T^{\qq, \kk, \pp}=0$.

Although the triad transfers contain the complete physical information related to the energetic coupling of scales, the sheer number of transfers involved makes them unusable in any direct manner. To ease our work, we decompose the spectral space into a series of structures and analyze the transfers that occur among them. The structures boundaries $s_\si \equiv(k_{\si-1},\ k_\si]$ are typically given as a power law in terms of the wavenumber $k$, $k_\si=k_0 \times 2^{(\si-1)/\Delta} $. The filtered ion distribution function $h^\si$ and filtered electric potential  $\bar\phi^\si$ are found to be:
\begin{eqnarray}
\{h, \bar \phi_1\}^\si(\kk)=\left\{ \begin{array}{lcl}
\{h, \bar \phi_1\}(\kk) , &  k \in s_\si \\
0 , &  k \notin s_\si
\end{array}  \right.. 
\end{eqnarray}
Depending on the selection of $k_0$ and $\Delta$, we obtain a $N$ number of wavenumber bands (or cylindrical like shells). Here, for $k_0=0.258$ and $\Delta=5$ we obtain $N=25$ shells. In real space, the total information can be recovered by summing over the inverse Fourier transform of each shell filtered contribution. 

The {\em triple shell transfer} occurring between the shell filtered quantities can be computed as,
\begin{align}
\mathcal S^{\si, \sii, \siii}&=\sum_{\qq\in s_\siii}\sum_{\pp\in s_\sii}\sum_{\kk\in s_\si} \mathcal T^{\kk, \pp, \qq}\delta_{\kk+\pp+\qq}   \;, 
\end{align}
and represent the basic information available to us for analysis. Knowing $\mathcal S^{\si, \sii, \siii}$ allows us to compute all other relevant nonlinear transfer quantities. By summing over all possible shells $\siii$ we can obtain the {\em shell-to-shell transfer} ($\mathcal P^{\si, \sii}$; implicitly defined below and analyzed previously \cite{BanonNavarro:2011p1274}) and the nonlinear transfer spectra by summing furthermore over $\sii$,
\begin{align}
& \mathcal T^\si=\sum_{\sii} \mathcal P^{\si,\sii}=\sum_{\sii}\sum_{\siii} \mathcal S^{\si,\sii,\siii} \;. \label{nettrans}
\end{align}
Numerically, when summing the transfer (\ref{nettrans}) over $\si$, which is equivalent to integrating the nonlinear transfer over the entire space, we obtain zero (comparable to machine precision). 

\begin{figure}[b]
\centering
\includegraphics[width = 1.0\columnwidth]{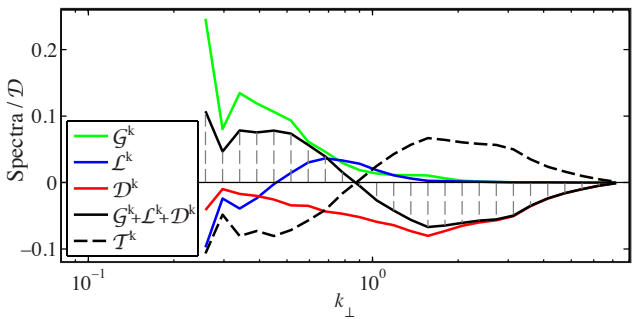}
\caption{The free energy rhs terms spectra normalized by the total dissipation rate $\mathcal D$. The vertical dashed lines represent the shell boundaries.}
\label{fig_linear}
\end{figure}

The spectral density contributions entering in the free energy balance equation, for perpendicular characteristic scales $k_\perp=k_K$, are presented in Fig.~\ref{fig_linear}. It is interesting to note that while the spectral density $\mathcal L_C^k$ is found to be zero, the $\mathcal L^k_\parallel$ term, although it integrates to zero globally, contributes to the overall linear term spectral form for time saturated states. This is important as the nonlinear transfer spectral density $\mathcal T^k$ is balanced by the sum of all the linear terms. The subsequent nonlinear transfers between scales can be seen as taking place under the constrained of a given transfer spectra. From this picture, the presence of the dissipation term at all scales is obvious.

\begin{figure}[t]
\centering
\includegraphics[width = 1.0\columnwidth]{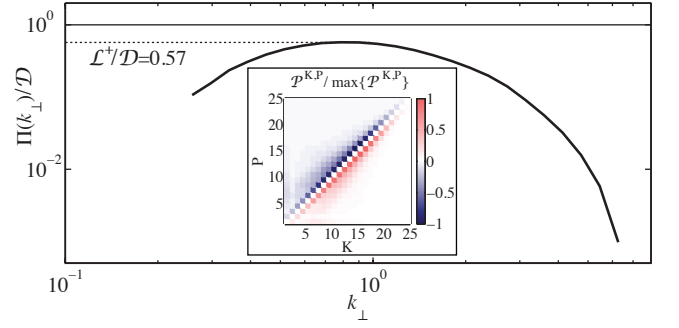}
\caption{The free energy flux across the shell boundaries normalized by the total dissipation rate $\mathcal D$. The insert picture depicts the shell-to-shell transfer for this run; for details see Ref.~\cite{BanonNavarro:2011p1274}.}
\label{fig_Flux}
\end{figure}

{\em Locality functions.---}The locality functions are defined from the triple transfers as a way to measure the non-locality degree of the triads which contribute to the energy scale flux. The flux through a scale (here, shell boundaries $k_\kc$) is defined by partial summing the transfer spectra $\mathcal T^\si$,
\begin{align}
\Pi(k_\kc)=\sum_{\si=\kc+1}^{N} \mathcal T^{\si}=\sum_{\si=\kc+1}^{N} \sum_{\siii=1}^{N} \sum_{\sii=1}^{N} \mathcal S^{\si,\sii,\siii}\;.
\end{align}
In Fig.~\ref{fig_Flux} we show the free energy flux across the perpendicular shell wavenumbers $k_\perp$. Since the source term contribution $\mathcal G^k$ is spread over a large interval, the flux across a scale $k_\perp$ builds up slowly to its cascade saturated value. Moreover, since the dissipation range is quite wide and permeates into the injection range, a true inertial range flux value can not be identified as the plateau on the flux, Fig~\ref{fig_Flux}. In fact, the scale flux plateau level is given by $\mathcal L^+$, representing the sum of the positive part of the linear contribution $\mathcal G^k + \mathcal L^k +\mathcal D^k$, here the first $10$ shells. The $\mathcal L^+/\mathcal D$ ratio clearly shows that only a fraction ($57\%$) of the energy injected into the system contributes to the nonlinear cascade.

Knowing the flux, the infrared (IR) locality function is defined by taking a probe wavenumber boundary $k_\kp$, so that $k_\kp \le k_\kc$, (sums considered in operatorial sense, as abbreviation, due to space limitations)
\begin{align}
\Pi_{\mbox{\scriptsize ir}}\,(k_\kp|k_\kc)=\sum_{\si=\kc+1}^{N}\left[\sum_{\sii=1}^{N}\right.& \sum_{\siii=1}^\kp \mathcal + \left.  \sum_{\sii=1}^{\kp}\sum_{\siii=\kp+1}^N  \right] \mathcal S^{\si,\sii,\siii} \;. \label{IRloc}
\end{align}
It measures the contribution to the flux through $k_\kc$ from triads of modes with at least one wavenumber less than $k_\kp$. In the second term, the sum over shell $\siii$ starts from $\kp+1$ to avoid double counting. In the limit $k_\kp \rightarrow k_\kc$, we recover the flux across the cutoff wavenumber $k_c$. It is customary to normalize the locality functions to the flux trough $k_c$, in which case a value of one is obtained for $k_p=k_c$ and less than one for $k_p/k_c<1$. Although the IR functions have a clear interpretation as the ratio of energy contributed to the flux through scale $k_\kc$ coming only from larger and larger scales, it should be remembered that for $k_p/k_c\ll1$ the transfers can only take place between the most non-local triads, i.e. triads with one wavevectors leg much smaller compared to the other two. Therefore, these functions can provide information regarding the locality of the nonlinear interaction. 

A similar definition is made for the ultraviolet (UV) locality functions, $k_\kc \le k_\kp$, (sums consider as for Eq.~(\ref{IRloc}), for brevity)
\begin{align}
\Pi_{\mbox{\scriptsize uv}}\,(k_\kp|k_\kc)=\sum_{\si=1}^{\kc}\left[\sum_{\sii=1}^{N}\right. &\sum_{\siii=\kp+1}^N+ \left.  \sum_{\sii=\kp+1}^{N}\sum_{\siii=1}^\kp \right] \mathcal S^{\si,\sii,\siii} \;,
\end{align}
which measures the contribution to the flux through $k_\kc$ from triads of modes with at least one wavenumber greater than $k_\kp$, therefore providing information regarding the locality makeup of a scale $k_\kc$ in relation with smaller and smaller scales.

\begin{figure}[tb]
\centering
\includegraphics[width = 1.0\columnwidth]{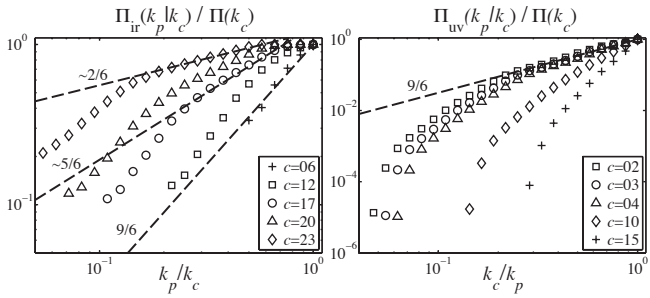}
\caption{The IR and UV locality functions, displayed for selected cutoff wavenumbers identified by the shell index $c$. Dashed lines equal or proportional to different power laws of the abscissae are displayed for reference.}
\label{fig_IR}
\end{figure}

Looking at the plot of ${\Pi_{\mbox{\scriptsize ir}}\,(k_p|k_c)}/{\Pi(k_c)}$ as a function of  $k_\kp/k_\kc$  and ${\Pi_{\mbox{\scriptsize uv}}\,(k_p|k_c)}/{\Pi(k_c)}$ as a function of  $k_\kc/k_\kp$ will reveal information related to the locality characteristic of the non-linear terms. 
The collapse of the locality functions dependence on $k_\kp$ for different values of $k_\kc$ represents a clear sign of self-similarity of the nonlinear interactions, which implies a dominance of the nonlinear terms in regard to the linear ones. Moreover, if the mentioned collapse exhibits a slope (in a log-log scale), then a state of asymptotic locality can be inferred, i.e. the nonlinear interactions saturated dynamically to the same level no mater how large the turbulence level becomes. From our simulations, Fig.~\ref{fig_IR}, none of these two behaviors can be clearly observed.

{\em Discussion and conclusions.---} Theoretically, an exponent for the IR and UV locality functions can be determined for an infinitely long inertial range. Considering the ${\mathbf v}_E\cdot\nabla h$ form of the nonlinearity, where ${\mathbf v}_E=\hat z \times \nabla \bar \phi_1$ and $\hat z\sim\nabla x\times\nabla y$, the triple shell transfer is found to have the form $S^{\si,\sii,\siii} \sim \langle h^\si({\mathbf v}^\siii_E\cdot\nabla) h^\sii \rangle$, where the angle brackets refers to volume averaging. Employing similar arguments as in Ref.~\cite{Aluie:2010p946} regarding the smoothness of scale filtered quantities, we determine a theoretical bound for the triple transfer as, $S^{\si,\sii,\siii} \le \langle |h^\si| \rangle \langle |{\mathbf v}^\siii_E| \rangle \langle |\nabla h^\sii| \rangle$. The scaling of the non-adiabatic part of the ion distribution function ($\langle |h^\si| \rangle \sim k_\si^{-1/6}$) and the $\mathbf E\times \mathbf B$ drift velocity scaling ($\langle |{\mathbf v}^\si_E| \rangle\sim k_\si^{-4/6}$) was given for two dimensional GK turbulence by Ref.~\cite{Plunk:2010p1360}, from the analysis of the self-similarity statistical symmetry of their respective increments. Using them, the triple scale transfer scales as, $S^{\si,\sii,\siii} \le q_\siii^{-4/6} p_\sii^{1-1/6} k_\si^{-1/6}$. Assuming a local dyadic transfer, i.e. $k_\siii=[k_\si/2-k_\sii, k_\si+k_\sii]$, in the IR limit ($k_\sii<k_\si/2$ therefore $k_\siii \approx k_\si$) or in the UV limit ($k_\sii>2k_\si$  therefore $k_\siii \approx k_\sii$) we obtain the scaling, $S^{\si,\sii,\siii} \sim p_\sii^{5/6} k_\si^{-5/6}$. This in turn translates as a $(k_\kp/k_\kc)^{\pm5/6}$ scaling for the IR and UV locality functions.

Although an asymptotic $5/6$ scaling of the IR and UV locality functions seams plausible and would indicate a more local interaction compared to magnetohydrodynamic turbulence ($2/3$, Refs.~\cite{Aluie:2010p946, Teaca:2011p1362}) but more non-local compared to fluid turbulence ($4/3$), these values can not be clearly identified from our simulations. First, we need to consider that the theoretical $5/6$ exponent is found in the limit of an infinite inertial range, an ansatz not verified in any range for GK turbulence. In spite of the local energy cascade Fig.~\ref{fig_Flux}-(insert), due to dissipation, the interaction of a given scale with smaller ones will be strongly damped increasing the scaling of the UV locality functions. The same scale will itself be damped compared to the larger scales, decreasing the IR locality function exponent.  

An effective non-local IR contribution signifies a dependence of GK turbulence on the type of instability driven it, while a stronger local UV depicts an insensitivity of GK's on the small scales and therefore the type of collision mechanism employed. Ultimately, these effects need to be properly accounted for in any numerical simulation.

{\em Acknowledgments.---} BT is supported by HP2C project, CSCS, Switzerland. ABN is supported by the contract of association EURATOM-Belgium state. Results achieved using computing resources of HPC-FF and JUGENE (through PRACE, including Tier-0), Julich, Germany.


\end{document}